# Deuterium-deuterium fusion charged particle detection using CR-39 and Deep Learning Model

Yuxing Wang [a], Allan Xi Chen [a*], Matthew Salazar [b], Nawar Abdalla [b], Zhifei Li [b], Benjamin Wrixon [b]

Alpha Ring International Limited

[a] 1631 W. 135th St. Gardena, CA 90249, U.S.A.
[b] 5 Harris Ct. Building B, Monterey, CA 93940, U.S.A.

**Abstract**

Columbia Resin #39 (CR-39) solid-state nuclear track detectors are widely used in fusion research for detecting charged particles produced in fusion reactions. However, analyzing increasingly complex and large-scale CR-39 track images to extract meaningful information is a tedious and time-consuming process, often prone to human error and bias. To address these challenges, we developed an AI-based classification model capable of differentiating protons, tritons, and helions produced during D-D fusion, using CR-39 track images as input data. The CR-39 track images were processed and used to train a deep learning model. By preprocessing the track images for noise reduction and feature enhancement, we trained the You Only Look Once version 8 (YOLOv8) network to distinguish the three particle types with high accuracy. We designed a method that cuts large images into smaller pieces and then enlarges them again to achieve a lower density in each segmented piece to improve accuracy. The sliced images are then reconstructed using overlapping cuts and Non-Maximum Suppression (NMS) to handle the overlapping parts of the stitches. By adopting NMS, we retain only the bounding box with the highest confidence level, avoiding duplicate detections and achieving accurate counting. The proposed model achieved a classification accuracy of over 94%, demonstrating its potential for improving automated track analysis in CR-39 detectors. Additionally, the model precisely identifies particle coordinates and counts, enabling comprehensive particle analysis. This study highlights the application of AI in track detection and classification, offering a robust solution for particle identification in CR-39 detector-based experiments.

## 1. Introduction

In fusion energy research, solid-state nuclear track detectors (SSNTD) like Columbia Resin #39 (CR-39) offer a robust and interference-resistant method for identifying charged particles from deuterium-deuterium (D-D) fusion reactions[1]. In many inertial confinement fusion experiments, CR-39 diagnostics are particularly valuable because they provide a means of detecting fusion products that are largely insensitive to electromagnetic interference (EMI) and X-rays, both of which are prevalent during such experiments [22]. D-D fusion is one of the most well-studied fusion reactions that produces multiple types of charged particles, such as protons, tritons, and helions. These charged products are key indicators for measuring overall fusion performance in various fusion experiments as well as characterizing nuclear interactions in beam-target accelerator systems. Measurements of the fusion particles are of particular importance, because they provide valuable information about plasma parameters, such as the rate of nuclear reactions occurring in plasma, the location of fusion-reaction sources, the role of thermal and accelerating mechanisms in fusion processes, etc[2]. Additionally, the D-D reaction is often preferred as a proxy for studying the D-T reaction, given the similar fuel chemistry of deuterium and tritium, but without the radioactive hazards associated with tritium [23]. Accordingly, precise quantification of the D–D fusion rate is essential for informing the design and optimization of future D–T fusion reactors. The 3.02 MeV protons leave the smallest tracks among the three particles due to their higher energy and lower mass. The tritons leave the largest tracks due to their slightly higher energy compared to the helions (1.01 MeV vs. 0.82 MeV). Additionally, the Triton track contrast is also much sharper by comparison. In this study, CR-39 detectors were used in conjunction with an Alpha Ring compact ion beam system (ARI-IBS) [3] to analyze particle tracks produced during D-D fusion. Figure 1 illustrates the experimental setup and image acquisition process, which includes chemical etching and high-resolution imaging using the TASL TASTRAK system--an advanced image acquisition system specifically designed for analyzing SSNTDs.

* Corresponding author e-mail address: allan@alpharing.com (Allan Xi Chen)

Previous studies using YOLOv8 have encountered challenges related to background impurities in the CR-39 processing workflow and the potential for repeated recognition of particles due to coupon partitioning. The research explored methods such as refining the particle labeling process and implementing data preprocessing techniques. However, a residual error rate of approximately 9% persisted[16]. To address the issue of particle detection along coupon slice edges, our study noted that due to the small and densely packed particles, multiple parameter adjustments were attempted but yielded suboptimal results. As a solution, the original large-scale images were segmented to ensure that small particles were represented with sufficient resolution, followed by scaling to the standard YOLOv8 input size of 640×640. This approach effectively reduced particle density within a single image. Furthermore, an overlapping tiling strategy was employed to prevent particles from being split across tiles. Overlapping regions were incorporated during the segmentation process to ensure that each particle was entirely captured in at least one tile. To mitigate duplicate detections in overlapping regions, the Intersection over Union (IoU) metric was utilized, with a threshold of 0.5. IoU is a measure of the overlap between the predicted and ground truth bounding boxes, calculated as the ratio of their intersection area to their union area, providing an evaluation of detection accuracy. Detections exceeding this threshold were merged into a single detection, with the bounding box having the highest confidence retained as the final detection. This methodology aimed to improve detection accuracy and ensure more reliable identification of particles in densely packed regions.

Our study focuses on utilizing these CR-39 tracks to train a deep learning model capable of particle classification, position annotation, and quantity estimation. This automated approach overcomes challenges posed by overlapping tracks and varying track characteristics, providing a scalable and efficient method.

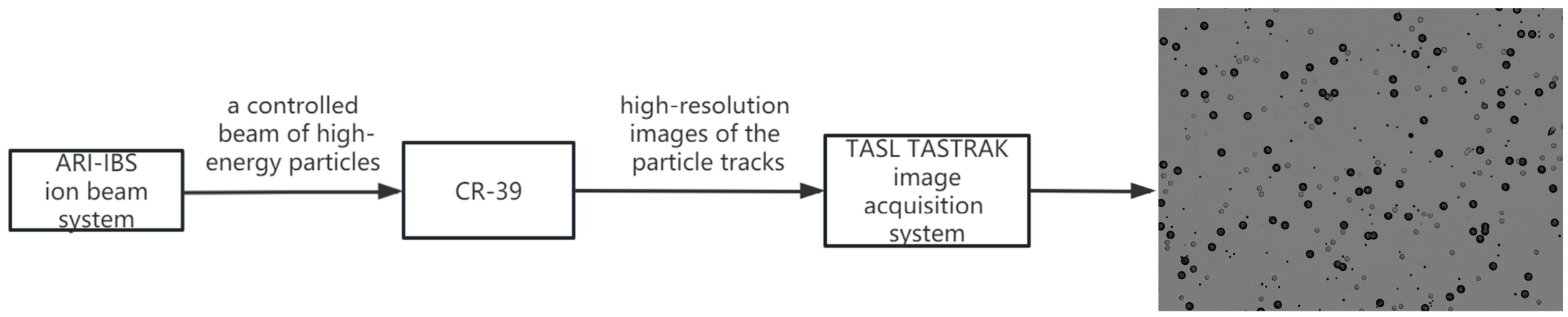

*Figure 1: CR-39 Particle Track Detection with ARI-IBS Ion Beam System and TASTRAK Imaging.*

## 2. Literature Review

### *2.1 Fusion charged particle detection*

The detection of charged fusion products plays a crucial role in nuclear fusion experiments. Various diagnostic techniques have been developed over the years. The cloud chamber allows direct observation of particle trajectories with the naked eye or via imaging systems [3]. In D-D fusion setups utilizing deuterated-diamond-like carbon (DDLC)-coated aluminum targets, a thin aluminum foil is often placed between the chamber and the detector to filter out low-energy particles. Only high-energy particles, like 3 MeV protons, can pass through and generate visible tracks, which appear as vapor streaks in the cloud chamber. In addition to visualizing the fusion reactions using the cloud chamber, we have also used CR-39 coupons for nuclear track detection. The filtering method using aluminum foils of varying thicknesses enables stepwise suppression of different charged particles—specifically, helions and tritons. At an aluminum thickness of 16 μm, only protons are energetic enough to leave visible tracks on the CR-39 or PM-355 detectors, effectively enabling selective proton detection[3]. In a quantitative approach, Szydlowski et al. utilized ion pinhole cameras equipped with PM-355 solid-state nuclear track detectors (SSNTDs) to measure the energy and angular distributions of fusion-generated protons. These detectors, placed behind calibrated pinholes, allow for spatial and directional resolution of charged particle fluxes. Their findings demonstrated that protons exceeding 3 MeV in energy could be effectively recorded and analyzed [4]. SSNTD is a well-established technique for detecting heavily ionizing charged particles. It records latent tracks in dielectric materials, which are later revealed through chemical etching and analyzed under microscopy. Due to its simplicity, low cost, and passive nature, SSNTD has become widely adopted. Pakistani researchers have refined etching methods, improved track registration in materials like CR-39 and mica, and applied the technique to study nuclear reactions, radon levels, and heavy ion interactions [5]. Although the etch pit sizes vary among the different particle types, overlap and proximity of these features often occur in the images, making it difficult to distinguish particles based on size or shape. Furthermore, while applying aluminum foils of different thicknesses can enhance particle separation by selectively blocking certain species, it also increases the complexity of particle classification. Therefore, a need exists for an automated, high-precision particle recognition technique capable of intelligently processing CR-39 etch pit images. Such a method should not only enable accurate classification of different particle types but also facilitate rapid and large-scale quantitative analysis, providing reliable support for measurements of nuclear reaction cross-sections, particle energy spectra estimation, and other relevant studies.

### *2.2 Introduction of Deep Learning*

Deep learning has demonstrated remarkable success in image-based detection tasks within real-world industrial environments, particularly under variable and challenging conditions. Deng et al. proposed a method utilizing a sparsely sampled rotating optical fiber array dosimeter combined with deep learning techniques to reconstruct two-dimensional dose distributions in radiotherapy. Their approach achieved a peak signal-to-noise ratio of 23.72, a root mean square error (RMSE) of 2.68, a structural similarity index (SSIM) of 0.95, and a gamma pass rate of 98.4% under the 3%/3 mm criterion[27]. Kim et al investigated the feasibility of deep learning (DL)-based dose estimation models for retrospective dosimetry. The study uses implemented outlier reduction techniques to address model inaccuracies arising from steep dose gradients. Overall, this research underscores the potential of DL techniques for fast and practical dose estimation in complex radiological scenarios, while outlining key directions for future optimization[29]. Wang et al. employed deep learning as the core technique for Fault Detection and Diagnosis (FDD) in Air Handling Units (AHUs). Due to the limited availability of labeled data, the study implemented a semi-supervised learning approach, where high-confidence pseudo-labeled data were combined with the original labeled dataset to enhance training effectiveness. This iterative self-training method significantly improved the model's classification performance. The final model achieved an overall F1-score of 85.6%[28]. Ken et al apply YOLOv3 to identify two types of etched patterns from merged microscope images of the detector's front and back surfaces. Etch pits representing incident (Nin) and non-reacted (Nout) particles were classified from CR-39 images. The accuracy demonstrates strong results with accuracies reaching 98.0 ± 4.0% for Nout and 97.3 ± 4.0% for Nin after 1000 training iterations[21]. Romeo et al. applied a deep learning approach based on Physics-Informed Neural Networks (PINNs) to model the physical process, and validated the model using experimental data, achieving a mean squared error of $2 \times 10^{-4} OD^2$. The method successfully overcomes limitations caused by diffusion effects, representing a significant effort to integrate artificial intelligence with real-world industrial scenarios[24]. Wang. applied eight types of image augmentation techniques to window detection and window state recognition tasks. The study highlights that the selection of augmentation strategies should be tailored to the specific architecture and use case—for example, brightness augmentation significantly improved AP50 in Faster R-CNN with MobileNetV3, while translation and hue augmentations proved most effective for YOLOv8 and R-FCN, respectively. These findings suggest that a context-aware combination of augmentation techniques can ultimately improve model accuracy across a wide range of applications[25]. What's more, Wang et al applied deep neural networks to classify building façade materials using Google Street View images. For London and Scotland, the model using MobileNetV3 with augmented data emerged as the most effective, achieving average accuracy rates of 75.65 % and 73.45%, respectively[6]. Similarly, Wang et al. conducted a comprehensive evaluation of eight augmentation methods—brightness, contrast, perspective, rotation, scale, shearing, translation, and a combined approach on tasks involving the classification of building stories and typologies using deep learning architectures. The study demonstrated that selecting the appropriate augmentation techniques in conjunction with the corresponding model can significantly enhance accuracy. For instance, applying brightness augmentation notably improved the classification accuracy of the "Residential Building" class, raising it from 96.77% to 99.17%[7]. And Park et al. explored the non-Protective Personal Equipment (PPE) detection applied the same set of eight augmentation techniques. The study applied these augmentations to a dataset of annotated site images across five PPE categories, and evaluated model performance using ViT, Swin Transformer, and PVT backbones. The article collectively underscores the practical value of data augmentation in improving deep learning performance, particularly when training data is limited or imbalanced[8]. And also, Dongjin et al. systematically explored the impact of eight distinct augmentation techniques on the deep learning model for the street view situation that some kinds of augmentation techniques can significantly enhance model accuracy across both detection and segmentation tasks [9]. G. Amit et al. have explored the use of deep learning techniques for particle energy classification on CR-39 detectors. The training of artificial neural networks (ANNs) enables the classification of individual tracks into discrete energy levels. As a result, data-driven machine learning models, which can extract complex and high-dimensional features from raw images, have become essential tools for achieving high classification accuracy[10]. Wang et al. proposed an automated rebar counting method using the Faster R-CNN deep learning model. By optimizing the model architecture and tuning hyperparameters, the researchers achieved high detection accuracy (94.61% AP50) and real-time processing speeds (up to 32 FPS). The results demonstrate that the deep learning method such as Faster R-CNN offer significant improvements in safety and productivity over traditional manual inspection methods[11]. Eum, I., *et al.* developed a YOLOv10-based object detection method for five categories of heavy equipment. By integrating advanced transformer-based backbones particularly the Swin Transformer- the model achieved high performance with metrics including AP50=89.72, AP50:95=66.71, IOU=84.81, APS=62.84, APM=86.39, and APL=92.53, which highlights the effectiveness of transformer-enhanced architectures in improving both accuracy and scalability for safety-critical applications in dynamic environments[26].

In industrial applications, YOLOv8[12][13] is extensively used to enhance the efficiency and accuracy of quality inspection, object detection, and classification tasks. Ye et al. have utilized an improved version of YOLOv8 for detecting objects in complex marine environments using side-scan sonars, addressing the challenges of target acquisition in such conditions[14]. [14]Additionally, Jie et al. have employed YOLOv8 for multi-scale object detection in synthetic aperture radar images, solving problem of the high probability of missed or incorrect detections[15]. From the above analysis, it is evident that deep learning model has broad applicability in industrial scenarios. By combining techniques such as data augmentation and model optimization, it can effectively address the challenges of object detection and classification in complex industrial environments. These results further demonstrate the practicality and effectiveness of YOLOv8 in CR-39 track image analysis, supporting its use in high-precision particle detection tasks.

## 3. Methods

The overall flowchart for data processing and model training for D-D fusion particle detection using a deep learning model is shown in figure 2 below. The object detection pipeline initiates with input processing, where raw data is preprocessed to optimize feature extraction. The process begins by splitting the images into smaller pieces. Subsequently, these images undergo various quality enhancement techniques, including denoising, contrast and brightness adjustment, and binarization. Following these improvements, image augmentation is performed to expand the dataset. After preparing the data, training parameters are set to effectively train the model. Once the model is trained, it is deployed for practical application. The final step involves testing the model with new images to evaluate its performance and obtain results.

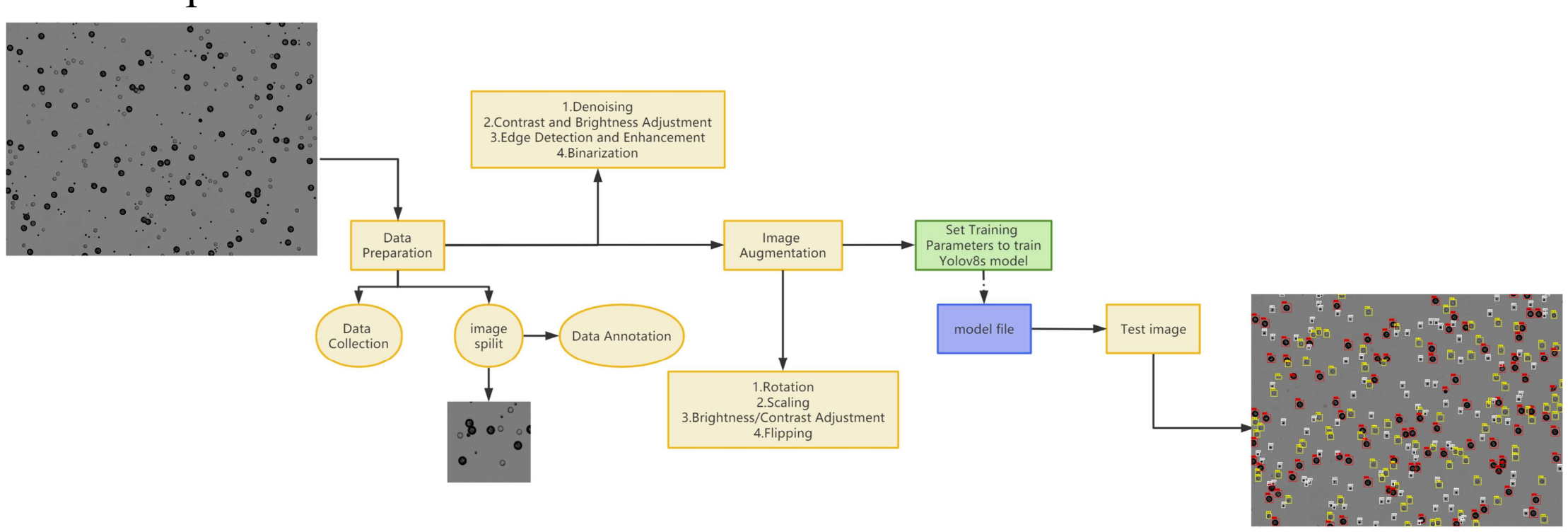

*Figure 2: Flowchart for data processing and model training for D-D fusion particle detection using deep learning model*

### *3.1. Data Processing*

High-resolution images of measurable nuclear tracks were obtained during D-D fusion experiments conducted under controlled conditions using the ARI-IBS. For training, 90 images with a resolution of 409×384 pixels were processed, while 10 images with the same resolution were reserved for validation. The images containing valid tracks of the three types of particles were confirmed by ARI experiment personnel and used to construct the dataset. Additionally, an independent test set consisted of 231 images with a resolution of 2048×1538 pixels. The raw images were divided into smaller blocks to reduce memory usage, improve computational efficiency, enhance local feature detection, minimize noise interference, and enable more detailed regional analysis. Each particle track was labeled as a proton, triton, or helion using the LabelImg tool to create the training and validation dataset. Subsequently, preprocessing steps, including denoising, contrast enhancement, and normalization, were applied to improve feature clarity and optimize the data for model training.

*3.2. Model Selection: YOLOv8*

In this study, the YOLOv8 model was employed for detecting and classifying CR-39 track images. The model can be divided into three main components as shown in figure 3 below: the Backbone, Neck, and Head. The model takes a 1×640×640 image as input and first passes it through the backbone for feature extraction. The backbone begins with a Conv2D layer using 32 filters and a 3×3 kernel with a stride of 2, followed by Batch Normalization and the SiLU activation function. This is followed by a series of C2f modules, which progressively increase the number of output channels and reduce the spatial resolution. A key innovation in the Backbone is the use of Cross Stage Partial (CSP) connections, which enhance the network's representational capacity and computational efficiency. CSP connections achieve this by splitting the feature map into two parts, processing one part through transformations while keeping the other unchanged, and then merging them, effectively reducing duplicate computation and enhancing the network's representation capability and inference speed[17]. The extracted features are then passed to the Neck, which incorporates a Path Aggregation Network (PANet) to perform multi-scale feature fusion and enhance the detection of small, medium, and large objects. PANet is an extension of the Feature Pyramid Network (FPN) that modifies the original top-down feature pyramid by introducing a bottom-up path augmentation, lateral connections, and adaptive pooling. The bottom-up path augmentation reconstructs feature hierarchies by adding an additional pathway that transfers low-level spatial details back to higher levels. Lateral connections integrate features between adjacent layers to maintain consistency across scales, while adaptive pooling helps aggregate global context information for more robust feature representation. Finally, the Head applies a detection layer to produce bounding box predictions and class probabilities for three target classes. It uses Batch Normalization and a Sigmoid activation function to generate the final outputs, consisting of N predicted bounding boxes. The detection Head adopts an anchor-free design, directly predicting object locations and classes, which improves flexibility and eliminates the need for predefined anchor boxes. Unlike anchor-based methods that rely on predefined box priors, anchor-free approaches predict object centers, offsets, and scales directly from feature maps, reducing computational complexity and enhancing adaptability to varying object scales and aspect ratios. In the post-processing stage, NMS eliminates redundant bounding boxes and filters predictions based on confidence scores, ensuring accurate and non-redundant detections. NMS is an algorithm used in object detection to retain the most relevant bounding box by suppressing those with high overlap (IoU) while keeping the one with the highest confidence score. YOLOv8's key innovations, including multi-scale feature fusion and an anchor-free design, significantly enhance its capability to detect complex and overlapping track structures across diverse scenarios. The model benefits from the parallel processing capabilities of the GPU, enabling real-time detection with optimized memory utilization and accelerated tensor operations through CUDA, making it highly efficient for large-scale data processing[18][19][20].

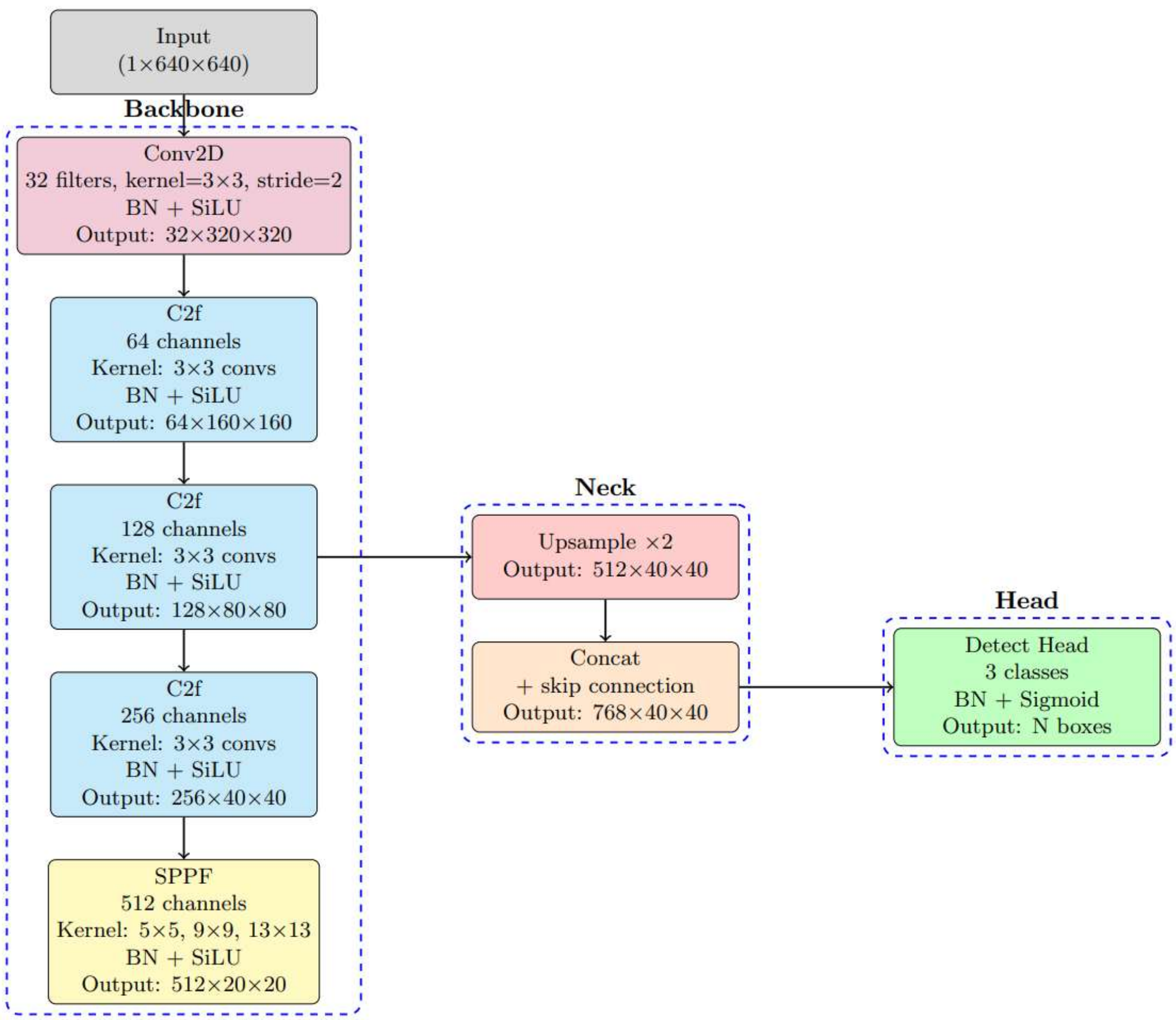

*Figure 3: Architecture of the YOLOv8-Based Object Detection Model for CR-39 Track Detection.*

### *3.3 Model Training*

In this study, the YOLOv8 model was employed for detecting and classifying CR-39 track images, utilizing an NVIDIA RTX A2000 GPU with 12 GB of memory to enhance computational efficiency and accelerate inference. The hardware and software configuration information is provided in Table 1 below.

*Table 1: Hardware and Software Configuration for AI training*

| **Hardware** | **Software** |
| --- | --- |

| | | | |
|---|---|---|---|
| GPU Model | NVIDIA RTX A2000 | NVIDIA-SMI Version | 552.23 |
| GPU Memory | 12 GB | Driver Version | 552.23 |
| Pytorch Version | 2.5.0 | CUDA Version | 12.4 |
| YOLO Version | v8s | Python Version | 3.12 |

To evaluate the model's performance, we used metrics including precision, recall, F1-score, and mean average precision (mAP) were utilized. These metrics comprehensively assess the accuracy of particle classification, the detection of overlapping particles, and the model's robustness under varying particle density conditions. The metrics are as follows:

$$\text{Precision} = \frac{\text{TP}}{\text{TP+FP}} \quad (1)$$

$$\text{Recall} = \frac{\text{TP}}{\text{TP+FN}} \quad (2)$$

$$\text{F1-Score} = 2 \cdot \frac{\text{Precision}\cdot\text{Recall}}{\text{Precision+Recall}} \quad (3)$$

$$\text{Mean Average Precision( mAP)} = \frac{1}{N}\sum_{i=1}^{N} \text{AP}_i \quad (4)$$

$$\text{AP} = \int_0^1 \text{Precision(Recall)}\ d(\text{Recall}) \quad (5)$$

In object detection evaluation, TP (True Positives) refers to instances where the model correctly identifies an object as belonging to the target class. FP (False Positives) occur when the model incorrectly identifies an object. FN (False Negatives) represent cases where the model fails to detect an object that actually exists, missing it entirely. Additionally, $AP_i$ (Average Precision for class $i$) quantifies the model's performance for a specific class by calculating the area under the Precision-Recall curve for that class. Precision measures the proportion of correctly identified positive samples among all predicted positives, while Recall reflects the model's ability to detect all relevant objects. F1-Score balances Precision and Recall, particularly for imbalanced classes. Average Precision (AP) represents the area under the Precision-Recall curve and Mean Average Precision (mAP) averages AP across all classes, providing a comprehensive measure of detection performance. Together, these metrics evaluate the model's accuracy, completeness, and reliability. The inference times averaged 361 ms per image.

### *3.4. Post-Training Validation*

After training, the model was rigorously evaluated on a separate set of CR-39 images that were not used during the training or validation phases. These test images consisted of data with varying particle densities, captured at different etching times (45, 90, and 180 minutes). As shown in the post training validation of the results (manual counting vs model counting) below, we assess the model's performance using manual counting by experts as a benchmark, ensuring accurate particle classification, precise localization, and count reliability. This process involved manually identifying and categorizing particles within each image, carefully comparing the model's predictions with the expert-verified counts. Particular attention was given to detecting overlapping tracks, accurately distinguishing between particle types, and verifying the total particle counts within each image. The manual annotations vs. AI model annotations for particle detection in microscopy are shown below (Table 2, Fig.4).

- *Table 2: Training and validation of results (manual counting vs. AI model counting)*

| | Manual Counting (5 images) | | | | | Model Counting (5 images) | | | | | Accuracy |
|---|---|---|---|---|---|---|---|---|---|---|---|
| image | 1 | 2 | 3 | 4 | 5 | 1 | 2 | 3 | 4 | 5 | total |
| triton | 92 | 93 | 83 | 103 | 106 | 93 | 94 | 87 | 106 | 108 | 98% |
| helion | 67 | 91 | 83 | 96 | 106 | 66 | 91 | 82 | 96 | 104 | 99% |
| proton | 85 | 105 | 107 | 78 | 90 | 87 | 108 | 109 | 80 | 94 | 97% |

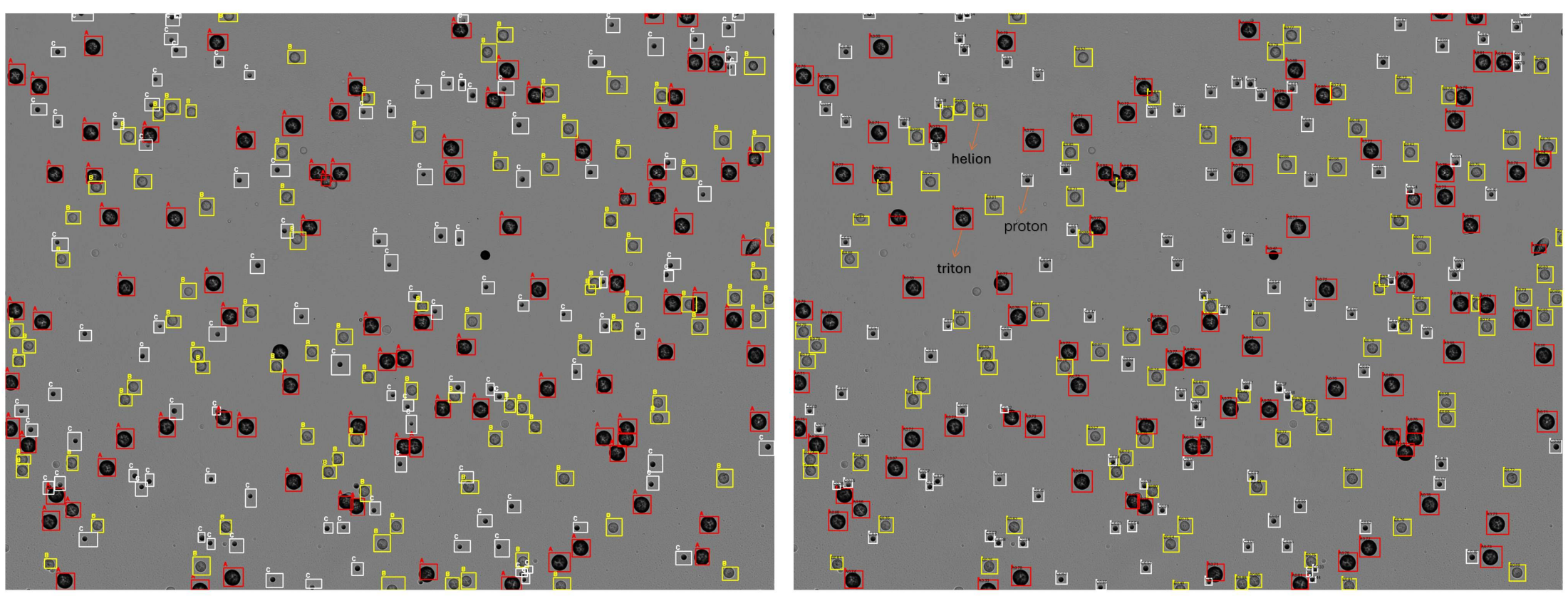

*Figure 4: Manual annotations of image2 (Left) vs. AI model annotations of image 2 (Right) for particle detection.*

### *3.5 Model Application*

The trained YOLOv8 model was integrated into a user-friendly graphical user interface (GUI) to facilitate automated analysis of CR-39 images (Fig. 5). The GUI is designed with an intuitive layout, divided into upper and lower sections. This interface supports single-image and batch processing, displaying labeled images alongside real-time particle counts. The original image is displayed on the left, while the processed image with labeled particles and bounding boxes is shown on the right. The interface also displays the quantities of each particle type and the total particle count in real time. When a folder is selected, images are processed one by one, and the outputs include processed/labeled image files and CSV summaries of particle classifications and coordinates. The CSV file contains the filename, particle classifications, bounding box coordinates for each particle, counts of each particle type, and the total particle count. The coordinates output by YOLOv8 represent the bounding box positions within the input image, measured in pixels. They include the top-left corner ($x_min, y_min$) and the bottom-right corner ($x_max, y_max$), defining the boundaries of the detected object. These coordinates are absolute values relative to the resized input image dimensions. The bounding box coordinates from the detection results are transformed back to the full-image coordinate system to ensure accurate placement on the original image, allowing data in overlapping regions to be overwritten for consistency. Subsequently, Non-Maximum Suppression (NMS) is applied to remove redundant detections in overlapping areas, merging all results and drawing the final bounding boxes on the original image. This process effectively resolves issues of missed detections at image boundaries, ensuring the completeness and accuracy of the detection results. For folder uploads, the CSV file also includes a summary of the total particle statistics for all processed images. This GUI system streamlines the workflow from image input to result output, enabling efficient and automated analysis of CR-39 images and significantly improving the accuracy and speed of the detection method.

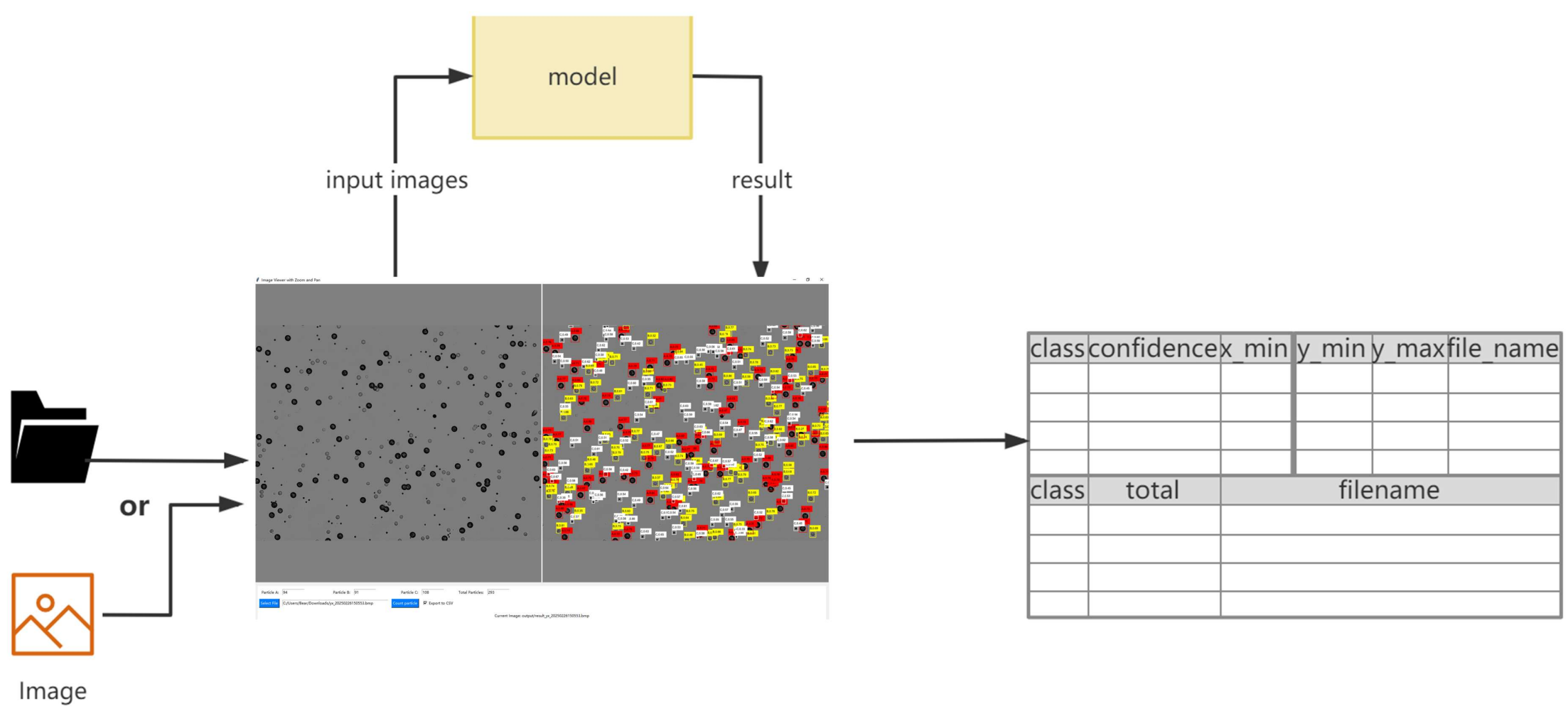

*Figure 5: CR39 Particle detection and classification pipeline: from image input to exporting the results.*

## 4 Results

The combined F1-score (Fig. 6) reaches its peak value of 0.91 at an optimal confidence threshold of 0.383, achieving a balance between precision and recall. This threshold provides robust performance across all particle classes, making it the most suitable choice for real-world applications in CR-39 particle detection and classification.

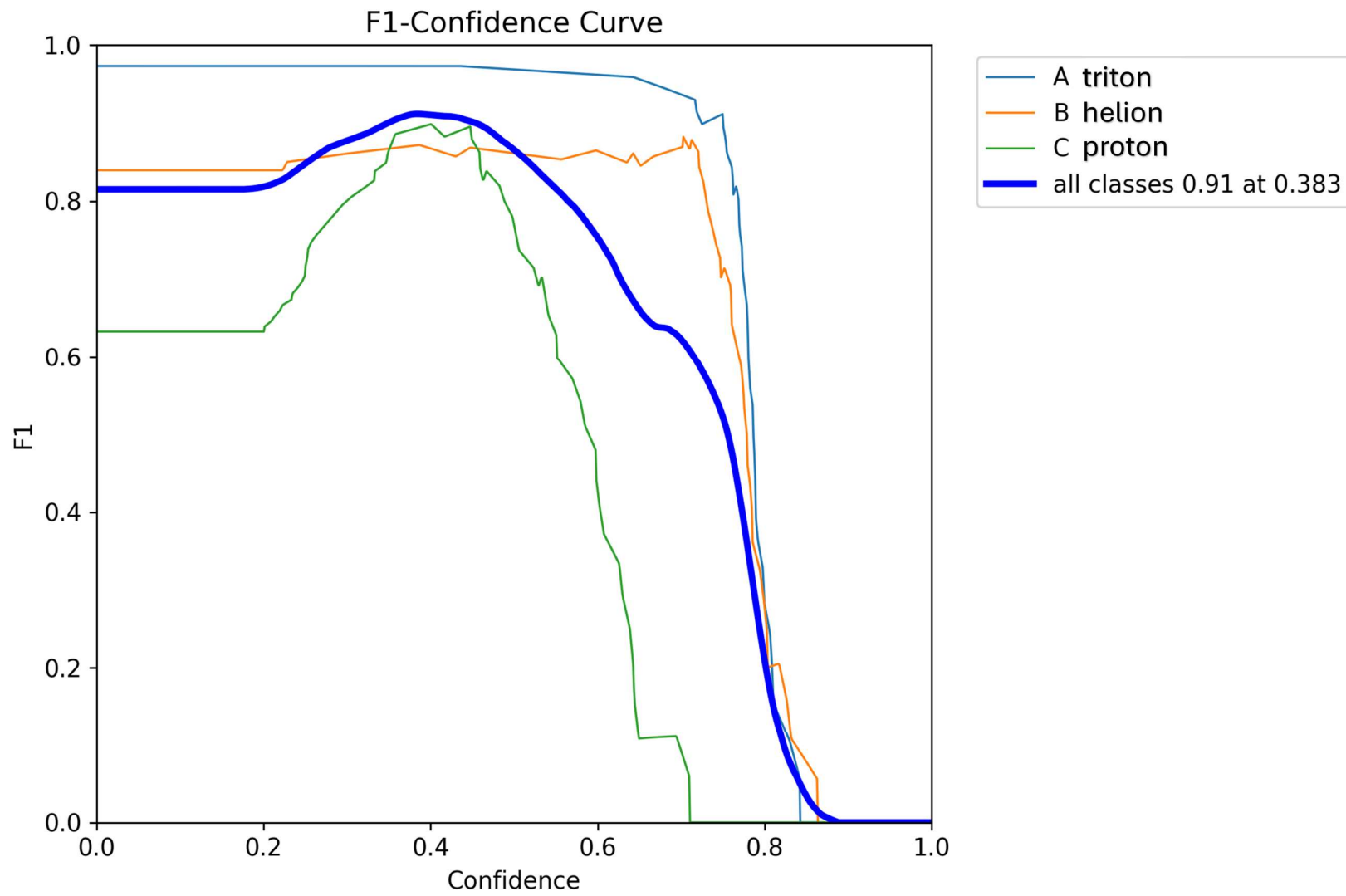

*Figure 6: F1-Confidence Curve*

The Precision-Confidence Curve (Fig. 7) shows that the triton class achieves the highest precision across most confidence thresholds, while the proton class exhibits greater fluctuation, indicating higher classification difficulty. The Precision-Recall Curve (Fig. 8) highlights the trade-off between precision and recall, with triton achieving the best performance (AP = 0.983), followed by helion (AP = 0.920) and proton (AP = 0.880). The overall mean average precision (mAP) for all classes is 0.928 at an IoU threshold of 0.5. The Recall-Confidence Curve (Fig. 9) demonstrates that the triton class maintains a high recall across a wide range of thresholds, while proton shows a significant drop in recall at higher confidence levels. These metrics collectively confirm the model's effectiveness in CR-39 particle detection and classification tasks.

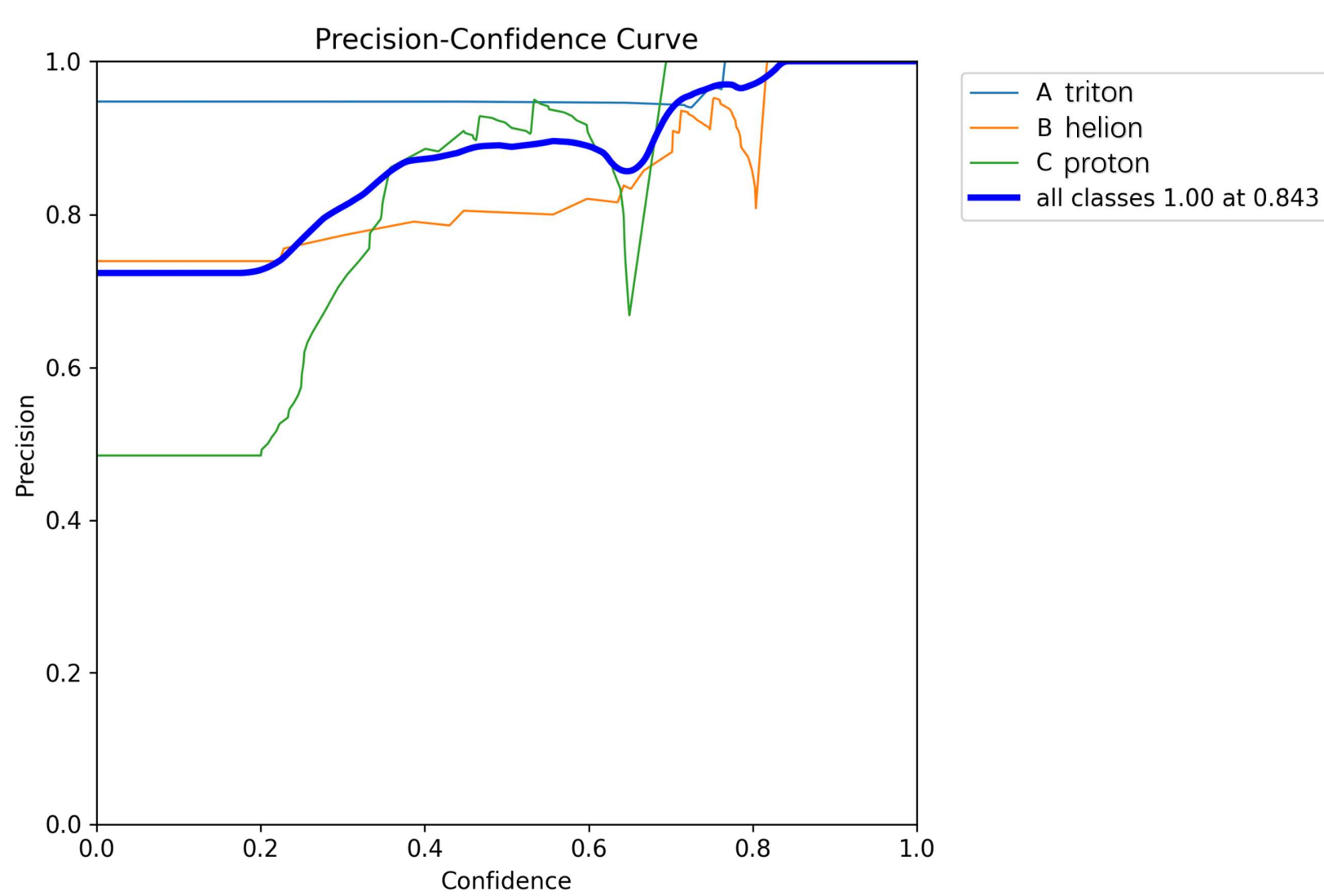

*Figure 7: Precision-Confidence Curve*

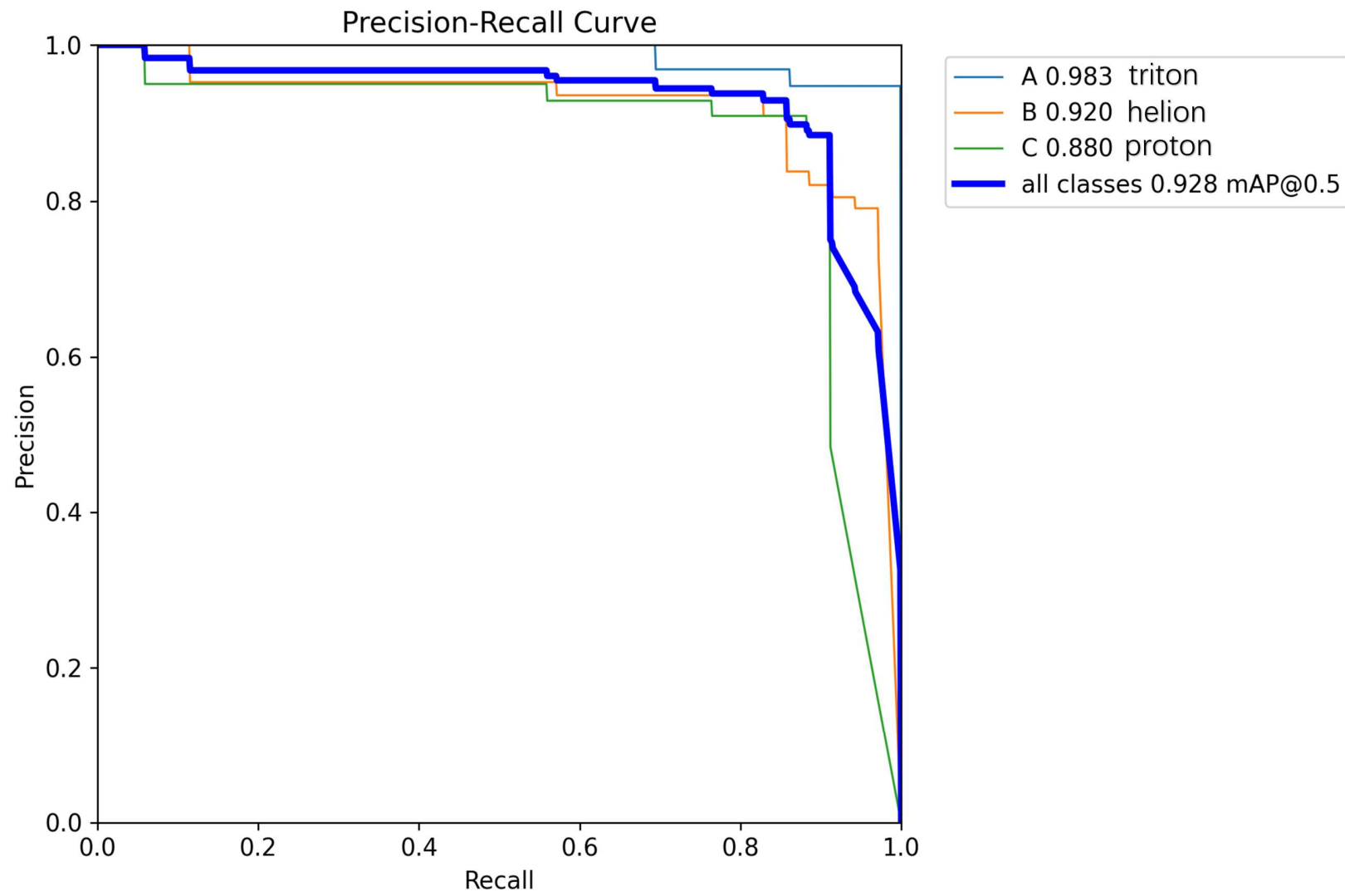

*Figure 8: Precision-Recall Curve*

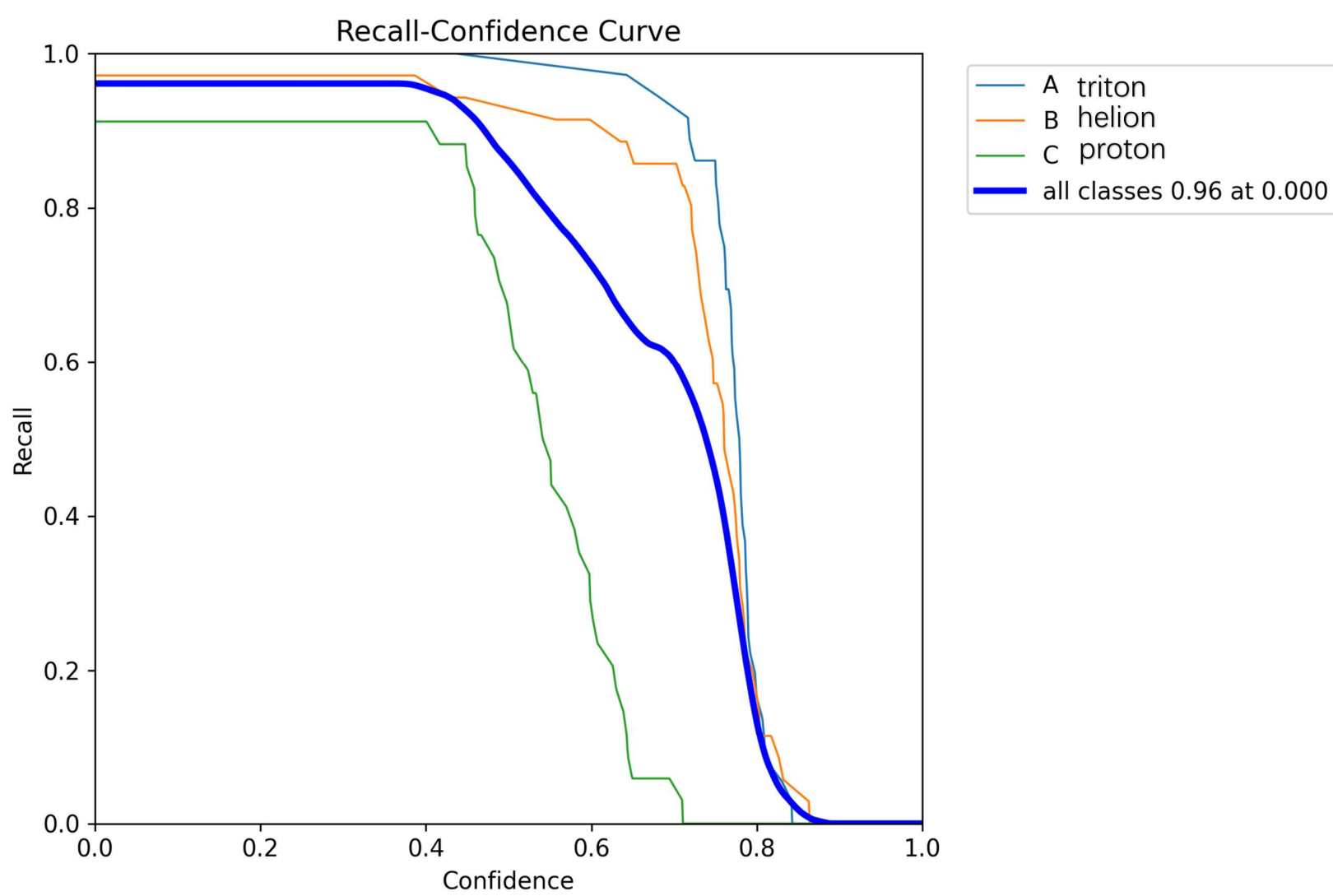

*Figure 9: Recall-Confidence Curve*

The training losses (including box loss, classification loss, DFL loss, and validation losses) are presented in figure 10. All loss curves show a consistent downward trend, demonstrating that the model is effectively learning, and that the optimization process is stable. For helions, the precision and recall metrics exhibit steady improvement throughout the training process, with both metrics nearing and exceeding 90%, indicating strong detection performance for this class. Additionally, the mAP50 and mAP50-95 metrics display an upward trend, confirming the model's ability to generalize across various IoU thresholds. The smooth convergence of the loss curves, coupled with the rising evaluation metrics, underscores the reliability of the model training process.

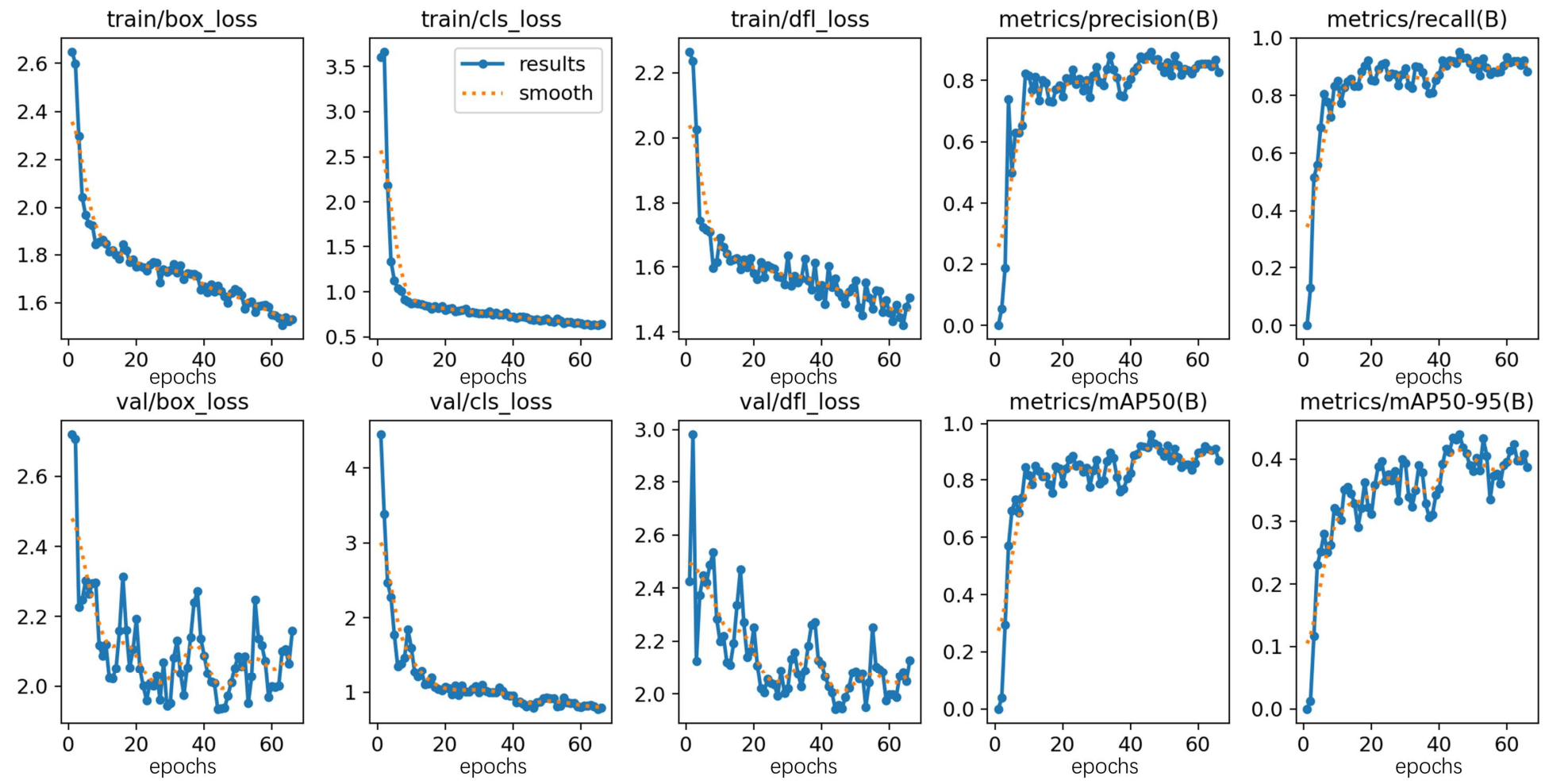

*Figure 10: Loss curves and evaluation metrics*

### *4.1 Background Noise*

To assess the model's robustness against background noise, we conducted detection experiments using 20 images containing only background noise as a control group as shown in Table 3. The results indicate that in 13 out of 20 images, the model did not detect any particles, demonstrating its effectiveness in filtering out background noise. However, in 7 images, the model incorrectly identified particles, misclassifying background impurities as actual particles. This suggests that while the model is generally effective in suppressing background noise, certain complex noise patterns may still lead to occasional false positives. Observations show that most noise particles are misclassified as helion particles. Since helion particles have distinct features, the model tends to predict them more frequently. Additionally, the similarity in shape between helion particles and certain background noise features leads to misclassification.

As illustrated in figure 11, both images belong to the background noise control group. The left image shows no detected particles, highlighting the model's capability to eliminate the influence of background noise. In contrast, the right image exhibits a single false positive detection, representing a minor misclassification case. These findings indicate that although the model demonstrates strong noise suppression, further improvements—such as enhanced data preprocessing and stricter confidence thresholds—may be necessary to minimize misidentifications caused by background artifacts.

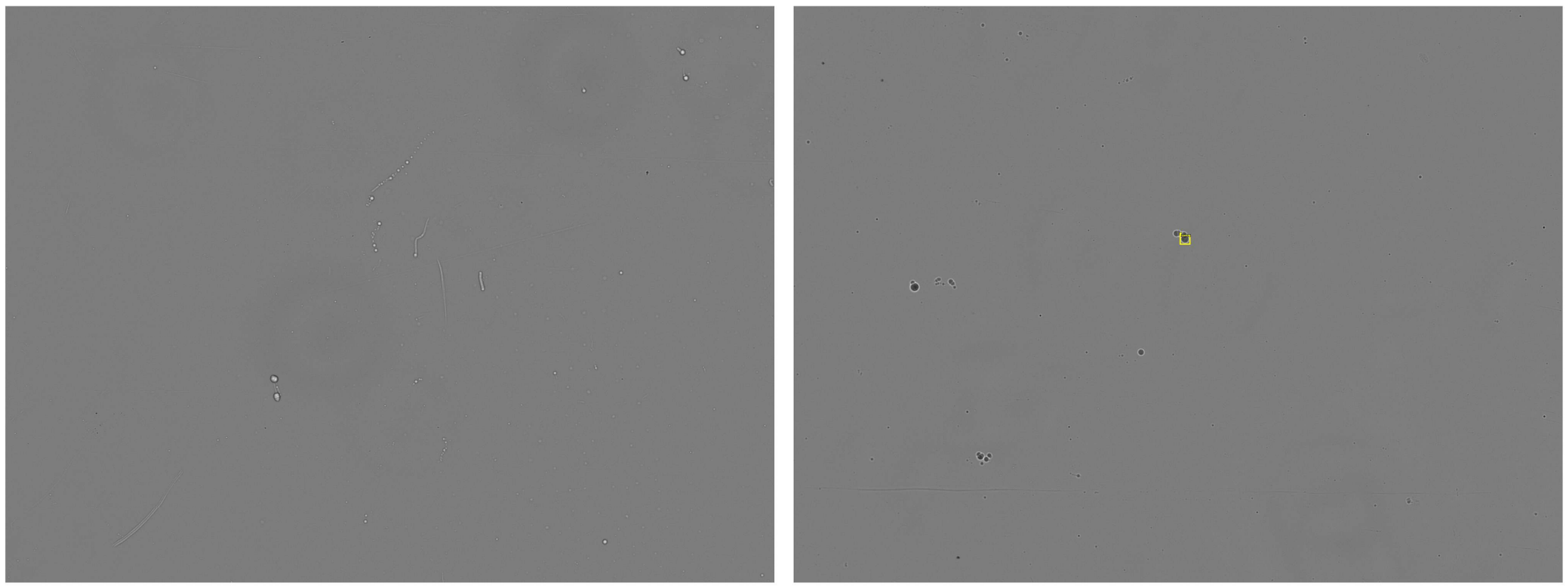

*Figure 11: Blank CR39 coupon for background noise control. (Left) Noise eliminated vs. (right) minor false detection.*

*Table 3: Sample detection results for 20 blank coupons to assess the background noise.*

| | Triton | Helion | Proton | #Total |
|---|---|---|---|---|
| 1 | 0 | 0 | 0 | 0 |
| 2 | 0 | 0 | 0 | 0 |
| 3 | 0 | 0 | 0 | 0 |
| 4 | 0 | 0 | 0 | 0 |
| 5 | 0 | 0 | 0 | 0 |
| 6 | 0 | 0 | 0 | 0 |
| 7 | 0 | 1 | 0 | 1 |
| 8 | 1 | 0 | 0 | 1 |
| 9 | 0 | 0 | 0 | 0 |
| 10 | 0 | 1 | 0 | 1 |
| 11 | 0 | 0 | 0 | 0 |
| 12 | 0 | 2 | 0 | 2 |
| 13 | 0 | 0 | 0 | 0 |
| 14 | 0 | 1 | 0 | 1 |
| 15 | 0 | 0 | 0 | 0 |
| 16 | 0 | 0 | 0 | 0 |
| 17 | 0 | 0 | 0 | 0 |
| 18 | 1 | 0 | 0 | 1 |
| 19 | 0 | 0 | 0 | 0 |
| 20 | 0 | 2 | 0 | 2 |

### *4.2 Comparison with ImageJ Analysis*

As shown in the Fig.12, left figure shows the traditional method using ImageJ for overlapping particles, each particle's center is marked. Where particles overlap, as with No.209 and No.235, the center is located at the intersection of the two particles, causing them to be recognized as a single particle. Additionally, while this method can identify the total number of particles, it is difficult to classify the particles based on their characteristics. The right figure utilizes the YOLOv8 model to recognize overlapping particles, and it can clearly distinguish them along their boundaries. The process involves pixel-to-physical size conversion, 8-bit grayscale conversion, and thresholding (0-100 pixel values as particles, the rest as background). The image is then converted into a binary mask with black background and white particles, followed by particle analysis that filters out excessively large, small, or non-circular particles to reduce misidentifications. This is because ImageJ typically uses a global threshold to segment the image into foreground and background. The pixel intensity in the overlapping area is similar to that of the area occupied by individual particles, making it difficult for the automatic thresholding algorithm to differentiate them as two separate entities. While ImageJ allows particle counting, it is limited in handling overlapping particles, often merging them into a single detection or filtering them out entirely, and is sensitive to background noise.

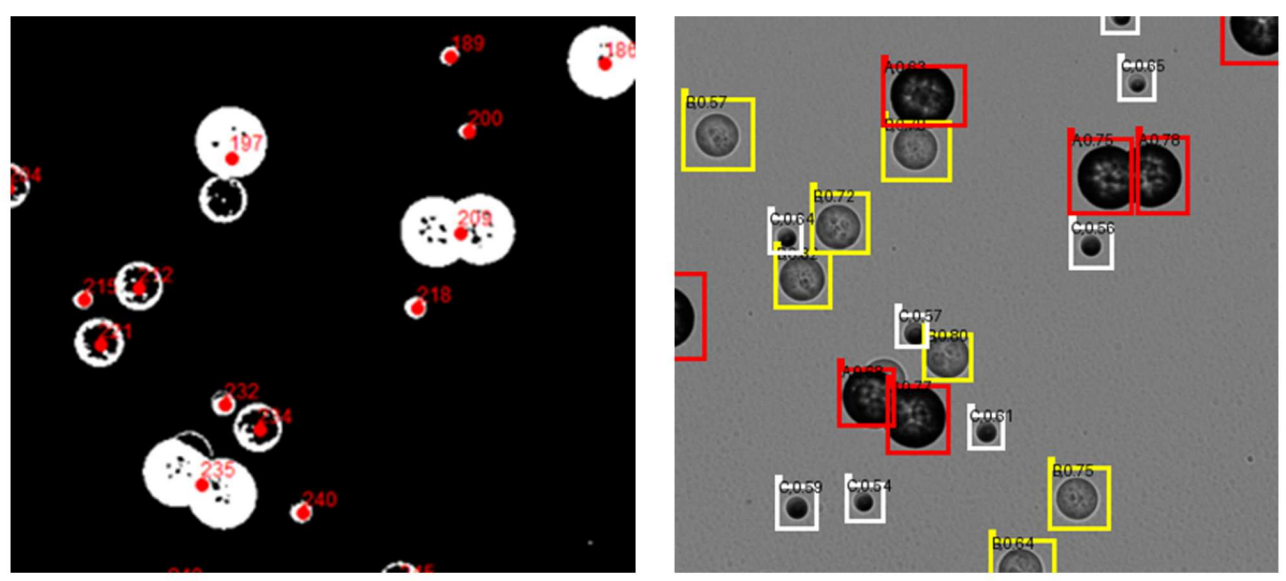

*Figure 12: Traditional image processing using ImageJ for overlapping particle recognition (left) vs . YOLOV8 model for overlapping particle recognition(right)*

YOLOv8, based on deep learning, overcomes some of these limitations by automatically adapting to different particle detectors and experimental conditions without manual parameter adjustments. It can handle variations in particle shape, size, density, and lighting conditions, making it more robust for complex detection tasks. Furthermore, GPU parallel processing significantly accelerates inference speed, achieving real-time detection in milliseconds. Unlike ImageJ, YOLOv8 eliminates the need for manual intervention, reducing subjective errors and enhancing both detection accuracy and efficiency, making it a more scalable solution for CR-39 particle analysis.

## 5 Conclusion

This study demonstrates that deep learning models, when combined with image splitting, composite techniques, and data augmentation, can effectively support object detection and classification in large-scale CR-39 track images, enabling the extraction of meaningful information. Additionally, the integration of a user-friendly interface significantly reduces analysis time and minimize human-induced bias. The trained YOLOv8 model demonstrates significant potential for real-world applications, particularly in automating CR-39 particle detection and classification processes. This is especially valuable for analyzing beam energy-dependent yields in D-D fusion. By correlating particle classification results with energy spectra, the model validates energy calibration and attenuation effects through accurate particle-type identification and localization. Its applicability also extends to STEM education, where integrating the model into automated analysis pipelines offers students a hands-on opportunity to efficiently interpret data, deepening their understanding of nuclear diagnostics and particle detection without the complexity of manual data processing.

Compared with other experimental methods—such as applying aluminum foils of varying thicknesses or using cloud chambers—our approach, based on deep learning and image processing, offers a more automated, scalable, and less labor-intensive alternative for CR-39 track analysis. In addition, compared with traditional image processing tools such as ImageJ, our method overcomes challenges related to overlapping track boundaries and the variability introduced by manually set threshold values. Relative to other deep learning methods, YOLOv8 is faster, more lightweight, more accurate, and better suited for detecting small or overlapping objects.

Despite the improvements achieved, the accuracy of detecting small particles still requires further enhancement, and interference from background noise remains a challenge. Future work will aim to address these issues by employing higher-resolution imaging systems, which can reduce background artifacts and improve the amount of useful information captured in each image. In addition, to further improve detection accuracy under diverse imaging conditions, future studies will explore architectural modifications to the FPN-PAN structure, integrate Deformable-LKA (Layer-wise Kernel Alignment) modules, and introduce a dedicated detection head for tiny particles, specifically designed to enhance recognition in high-density and overlapping scenarios.

## 6 Acknowledgement

The authors would like to thank Prof. Roger Falcone, Prof. Richard Petrasso, Prof. Jyhpyng Wang, Dr. Kosta Yanev, Dr. Nathan Eschbach, Mr. Iskren Vankov, Mr. Christian Yoo, Ms. Qiong Wang, Mr. Peter Liu, Mr. Paul Chau, Ms. Belinda Mei, Dr. Peter Hsieh, Mr. Wilson Wu, and Mr. Charles Wu for valuable discussion and insight. This work was conducted by scientists and engineers as employees of Alpha Ring International Limited and its affiliated companies. This study was funded by Alpha Ring International Limited, which supports the education and training of a fusion industry workforce. The funding from Alpha Ring International Limited did not influence the scientific integrity or the results of this study.